\documentclass[a4paper, 12pt]{amsart}
\usepackage[utf8]{inputenc}

\usepackage{amssymb, amsmath}
\usepackage{amsfonts}
\usepackage{float}
\usepackage{graphics}
\usepackage{subfigure}
\usepackage{color}
\usepackage{graphicx}
\usepackage{enumitem}
\usepackage[unicode, colorlinks, pdfpagelabels]{hyperref}
\usepackage{tikz}
\usepackage{pgfplots}
\pgfplotsset{compat=newest}

\usetikzlibrary{decorations.markings}
\tikzset{->-/.style={decoration={
  markings,
  mark=at position #1 with {\arrow{latex}}},postaction={decorate}}}
\tikzset{
    set arrow inside/.code={\pgfqkeys{/tikz/arrow inside}{#1}},
    set arrow inside={end/.initial=>, opt/.initial=},
    /pgf/decoration/Mark/.style={
        mark/.expanded=at position #1 with
        {
            \noexpand\arrow[\pgfkeysvalueof{/tikz/arrow inside/opt}]{\pgfkeysvalueof{/tikz/arrow inside/end}}
        }
    },
    arrow inside/.style 2 args={
        set arrow inside={#1},
        postaction={
            decorate,decoration={
                markings,Mark/.list={#2}
            }
        }
    },
}
\graphicspath{{./figures/}}

\newcommand{\abs}[1]{\lvert #1 \rvert}

\newcommand{\norm}[1]{\| #1 \|}

\newtheorem{theorem}{Theorem}[section]

\title[Dynamical Transition Theory of Hexagonal Patterns]{Dynamical Transition
  Theory of Hexagonal Pattern Formations}
\author{Taylan Şengül}
\address{Department of Mathematics, Marmara University, 34722 Istanbul, Turkey}
\email{taylan.sengul@marmara.edu.tr}
\date{\today}

\begin{document}
\maketitle
\begin{abstract}
  The main goal of this paper is to understand the formation of hexagonal
  patterns from the dynamical transition theory point of view. We consider the
  transitions from a steady state of an abstract nonlinear dissipative system.
  To shed light on the formation of mixed mode patterns such as the hexagonal
  pattern, we consider the case where the linearized operator of the system has
  two critical real eigenvalues, at a critical value $\lambda_c$ of a control
  parameter $\lambda$ with associated eigenmodes having a roll and rectangular
  pattern. By using center manifold reduction, we obtain the reduced equations
  of the system near the critical transition value $\lambda_c$. By a through
  analysis of these equations, we fully characterize all possible transition
  scenarios when the coefficients of the quadratic part of the reduced equations
  do not vanish. We consider three problems, two variants of the 2D
  Swift-Hohenberg equation and the 3D surface tension driven convection, to
  demonstrate that all the main theoretical results we obtain here are indeed
  realizable.
\end{abstract}

\section{Introduction: Main assumptions and results}
Transition phenomena is throughout all nonlinear sciences \cite{ptd,
  kuznetsov2004, hoyle2006}. It shapes many physical, biological and social
systems through instabilities. The formation of patterns in such systems,
whether it be coatings of animals \cite{murray2003}, convection cells in fluid
systems \cite{benard-1901} or crime patterns in cities \cite{short2010}, is
intrinsically related to the transitions taking place in those systems. One of
the tools to understand and classify the transition behavior is the dynamic
transition theory \cite{ptd}. The current work is an attempt to combine this
theory with certain aspects of pattern formations and relies on some of the
previous work in this direction \cite{dijkstra2013, sengul2013, sengul2014,
  liu2015}.

\subsection{The setting and the main assumptions}
\label{sec: setting}

We are interested in the transitions of a steady state solution of a general
nonlinear dissipative system \cite{temam1997} on a Hilbert space $X$
\begin{equation}\label{main}
  \frac{du}{dt} = L_{\lambda}u + G_{\lambda}(u), \quad t>0
\end{equation}
where $u: [0,\infty) \mapsto X$ is the unknown function and $\lambda \in
\mathbb{R}^1$ is a parameter. Here $L_{\lambda}: X_1 \to X$ is a linear operator
where $X_1$ is another Banach space with compact and dense inclusion $X_1
\subset X$ and $G_{\lambda}$ is a nonlinear operator satisfying certain
properties given later.

\subsubsection{The assumptions on the spectrum of the linear operator}

We will assume that the linear operator $L_{\lambda}$ has a countably infinite set of eigenvalues
\begin{displaymath}
  \{ \beta_i(\lambda) \in \mathbb{C} : i \in \mathbb{N} \}
\end{displaymath}
with a complete set of eigenvectors
\[
  \{f_i \in X_1, i=1,2,\dots\}
\]
satisfying the following conditions on its spectrum, known as the \textbf{PES
  conditions}.
\begin{equation} \label{PES}
  \begin{aligned}
    & \beta_1(\lambda),\beta_2(\lambda)\in \mathbb{R},\\
    & \beta(\lambda):=\beta _1(\lambda)=\beta_2(\lambda) 
      \begin{cases}
        <0 & \lambda <\lambda _{c} \\
        =0 & \lambda =\lambda _{c} \\
        >0 & \lambda >\lambda _{c}
      \end{cases} \\
    & Re \beta _i<0,\qquad \forall i=3,4,\dots
  \end{aligned}
\end{equation}
Much of the linear theory on stability and transitions involves establishing the
PES conditions, see \cite{chandrasekhar1961} for the classical fluid dynamics and
\cite{pedlosky1987} for the geophysical fluid dynamics.

\subsubsection{The assumptions on the physical space}
As the physical space, we assume a bounded spatial domain with at least two
dimensions. We also distinguish between the eigenvectors $f_i$ of the linear
operator and a (possibly) distinct set of basis vectors
\begin{displaymath}
  \{ e_{i_1, i_2}^j \in X_1 : i_1, i_2, j \in \mathbb{N} \}
\end{displaymath}
indexed by the wave indices $i_1$, $i_2$ spanning the two horizontal spatial
dimensions and $j$ is the index of the other directions which we usually
suppress for ease of notation. Moreover, we assume that the first two critical
modes have the spatial structure
\begin{equation} \label{eigenmode-assumption}
  f_1 \sim e_{j_1, j_2}, \qquad
  f_2 \sim e_{k_1, 0},
\end{equation}
for some non-negative integers $j_1, j_2, k_1$. Even for 2D problems, the basis
vectors $e_{i, j}$ are usually different from the basis vectors $f_j$ of the
linear operator. \textit{For example}, in a scalar reaction diffusion type
equation, the vectors $e_{i, j}$ are usually the eigenbasis of the Laplacian
operator with the given boundary conditions, while for fluid problems, the
vectors $e_{i, j}$ are the eigenbasis of the Stokes operator, as we consider in
\autoref{sec: marangoni convection}.

\emph{As an example}, if the spatial domain of interest is a rectangular domain
$(0, L_1 \pi) \times (0, L_2 \pi) \times (0, 1)$ with Neumann boundary
conditions in the horizontal directions, then the eigenmodes $f_j$ are given by
\begin{displaymath}
  f_j = \sum_{k=1}^{\infty} \hat{f}_k e_{j_1, j_2}^j(x, y, z)
\end{displaymath}
where
\begin{equation} \label{cosine-modes}
  e_{j_1, j_2}^j = \cos (j_1 x_1 /L_1) \cos (j_2 x_2 /L_2) w_j(z)
\end{equation}
where $j_1$, $i_2$ are non-negative integers and $w_j$ are basis functions
satisfying the vertical boundary conditions.

\begin{figure}[H]
  \centering \includegraphics[scale=.4]{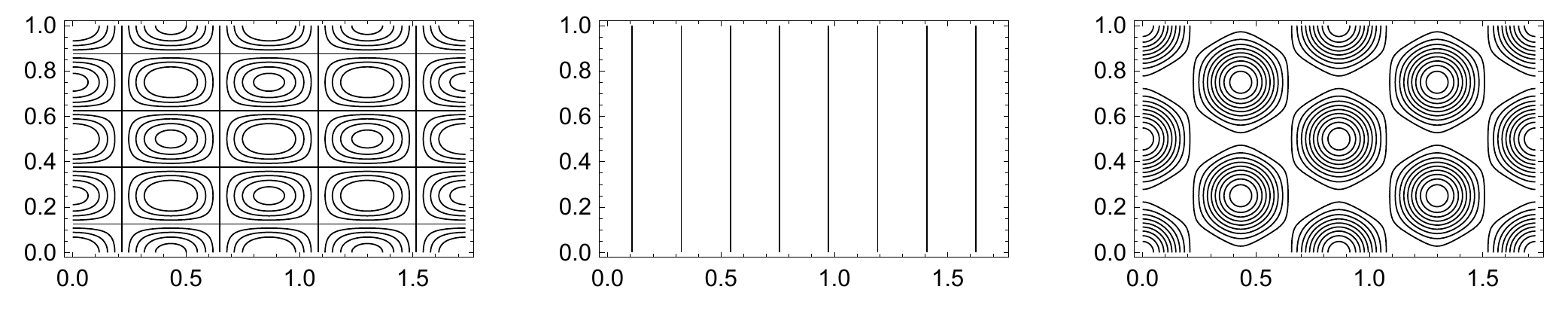}
  \caption{ (a) A rectangular mode $\cos (\frac{ 4 x_1 }{ \sqrt{3} }) \cos (4
    x_2)$ (b) A roll mode $\cos (\frac{ 8 x_1 }{ \sqrt{3} }) $ (c) A mixed mode
    $4 \cos (\frac{ 4 x_1 }{ \sqrt{3} }) \cos (4 x_2) + \cos (\frac{ 8 x_1 }{
      \sqrt{3} }) $. }
  \label{fig:rec-rollmode}
\end{figure}

Under the assumption \eqref{eigenmode-assumption}, the eigenmode $f_1$
represents a rectangular (horizontal) pattern, and the eigenmode $f_2$
represents a roll pattern. Their linear combinations give rise to mixed patterns
such as the hexagonal pattern, see \autoref{fig:rec-rollmode}.

\begin{figure}[H]
  \centering
  \includegraphics[scale=.3]{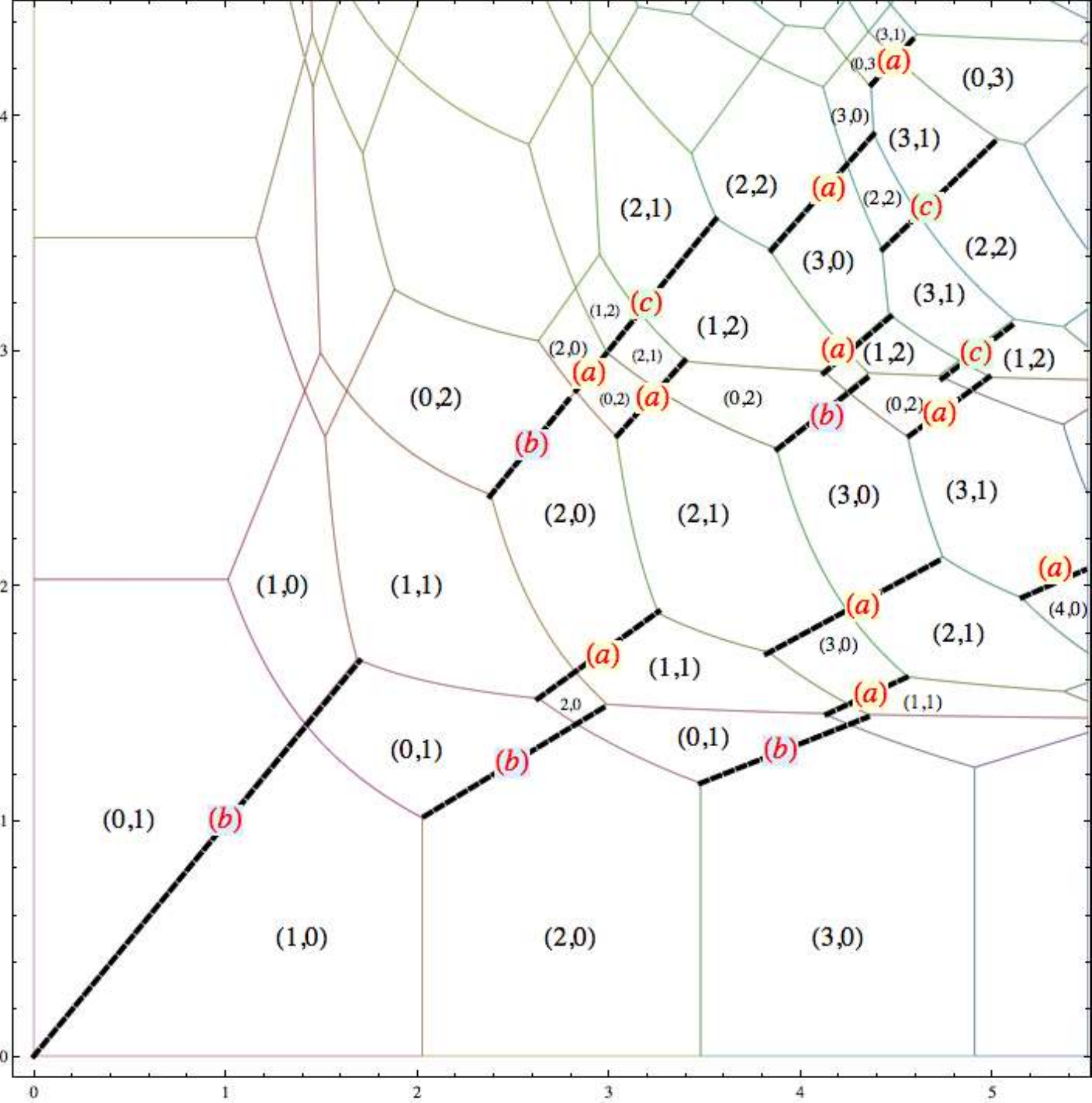}
  \caption{The critical index map for the 3D Rayleigh-Benard convection with
    respect to the horizontal length scales $L_1$ and $L_2$, from
    \cite{sengul2013}. }
  \label{fig:wavemap}
\end{figure}

In many physical examples, the critical modes are selected according to the
horizontal wave number, that is the first two critical modes have equal
horizontal wave numbers. This implies,
\eqref{eigenmode-assumption} means
\begin{equation} \label{wave-no}
  \left( \frac{ j_1 \pi }{ L_1 } \right)^2 + \left( \frac{ j_2 \pi }{ L_2 }
  \right)^2 = \left( \frac{ k_1 \pi }{ L_1 } \right)^2 
\end{equation}
from which $k_1 > j_1$ follows. We remark here that, the equality of the wave
numbers of the first two critical modes, imposes a severe relation on the
horizontal aspect ratio $L_2/L_1$ of a rectangular domain.
\begin{equation} \label{L1L2rel}
  L_1 = \sqrt{ \frac{  k_1^2 - j_1^2  }{ j_2^2 } } L_2.
\end{equation}
As a result, in applications, this type of transition is non-generic, that is,
does not occur if the aspect ratio is chosen randomly. We give the choice of the
wave indices for the 3D Rayleigh-Benard convection with free slip boundary
conditions in \autoref{fig:wavemap}. The figure shows the non-genericity of the
higher multiplicity transitions and the length scales at which double equal wave
number mode transitions occur for (a) a roll and a rectangle mode, (b) two roll
modes, (c) two rectangle modes.

\subsubsection{The main assumptions on the nonlinear operator}

We assume that $G$ consists of higher order terms in $u$, that is
$G_{\lambda}(u) = o(\norm{ u }_{X_{\alpha}})$ where $X_{\alpha}$ is an
interpolation space with $0 \le \alpha < 1$. This ensures that \eqref{main}
admits the homogeneous steady state solution
\begin{displaymath}
  u(t) = 0, \qquad \forall t \ge 0.
\end{displaymath}

We consider the following Taylor expansion of $G$.
\begin{equation} \label{Taylor}
  G(u) = G_2(u, u) + G_3(u, u, u) +  \cdots
\end{equation}
Here $G_2$ is the bilinear and $G_3$ is the trilinear operator of the Taylor
expansion of $G$ and the rest of the expansion will not play a role in the
analysis. 

\textit{The main assumption} is the following orthogonality conditions on the bilinear
$G_2$ and trilinear $G_3$ parts of the nonlinear operator with respect the basis
vectors $e_{i, j}$. We assume that if $\pm i_r \pm j_r \ne \pm k_r$ for some
choice of $\pm$ and at least one of $r= 1, 2$ then
\begin{equation} \label{main-nonlinear-assumption}
  \langle G_2(e_{i_1, i_2}, e_{j_1, j_2}), e_{k_1, k_2} \rangle = 0 .
\end{equation}
Here $\langle \cdot \rangle$ represent the inner product of $X$.
Similarly, $\pm i_r \pm j_r \pm k_r \ne \pm l_r$ for some choice of $\pm$ and at
least one of $r=1, 2$, we assume that
\begin{equation} \label{main-nonlinear-assumption2}
  \langle G_3(e_{i_1, i_2}, e_{j_1, j_2}, e_{k_1, k_3}), e_{l_1, l_2} \rangle = 0 .
\end{equation}
Such orthogonality conditions are typical for trigonometric basis functions and
nonlinear operators which are products of functions and their derivatives. Our
main assumptions are satisfied in many physically interesting systems such as
the convective motions of fluids \cite{lappa2009, sengul2014, sengul2013,
  dijkstra2013, han2018}, reaction-diffusion systems \cite{murray2002, li2016,
  ong2016, zhang2018} and pattern formation equations \cite{cross2009, yang2018,
  cross1993}. We will also present several applications where these assumptions
hold in \autoref{sec:Applications}.

For example for modes given by \eqref{cosine-modes} and a general nonlinear
operator of the form
\begin{displaymath}
  G(u) = a_1 u^2 + a_2 u u_x + a_3 u u_y + a_4 u^3 + a_5 u^2 u_x + \cdots
\end{displaymath}
where $a_i$ are constants and the usual $L_2$ inner product, the assertions hold
true due to the orthogonality of trigonometric functions.

\subsection{Discussion of the main results}

We first derive the general structure of the reduced (amplitude) equations by
using the center manifold reduction. Letting $u_1(t) f_1 + u_2(t) f_2$ to denote
the center part of the solution, we obtain the following equations.
\begin{displaymath}
\begin{aligned}
  &\frac{du_1}{dt}=\beta(\lambda) u_1 + a_1 u_1 u_2
  && + u_1 (a_2 u_1^2+a_3 u_2^2)+ O(4), \\
  &\frac{du_2}{dt}=\beta(\lambda) u_2 + b_1 u_1^2
  && + u_2 (b_2 u_1^2+b_3 u_2^2)+ O(4).
\end{aligned}
\end{displaymath}
These equations describe the long time behaviour of the system, near the
transition point $\lambda = \lambda_c$ close to the basic steady state solution.
The reduced equations consist of a quadratic part with coefficients $a_1$, $b_1$
due to the bilinear interactions between the critical modes and a cubic part
with coefficients $a_2$, $a_3$, $b_2$ and $b_3$ due to the bilinear interactions
of the critical modes with the higher frequency modes plus trilinear self
interactions of the critical modes. Our analysis of the reduced equations shows
that when none of the coefficients $a_1$, $b_1$, $b_3$ vanish, the type of
transition depends only on these three parameters.

In this paper, we address the case $a_1 \ne 0$, $b_1 \ne 0$ and $b_3 \ne 0$. In
the case $a_1 = b_1 = 0$, there are no bilinear interactions among the critical
modes, and the behavior of the system is determined by the cubic coefficients
$a_2, a_3$, $b_2$, $b_3$. That case is generic case when the first two critical
modes are both roll-type or both rectangle-type and is also often encountered in
the applications \cite{sengul2013}. It also occurs under certain symmetry
conditions which frequently arise in nonlinear systems of interest. We will
address this case in a future study.

Next, by a detailed analysis of the reduced equations, we describe the
bifurcated steady states and their stability and describe all the possible
transition scenarios. Due to the interactions of these two modes, a variety of
new states emerge after transition, including those associated with hexagonal
patterns.

In terms of transition analysis, our guiding principle is the dynamic transition
theory of Ma and Wang \cite{ptd}. The key philosophy of dynamic transition
theory is to search for the full set of transition states, giving a complete
characterization of stability and transition. The set of transition states is a
local attractor, representing the physical reality after the transition. As a
general principle, dynamic transitions of all dissipative systems are classified
into three categories: \emph{continuous (Type-I), catastrophic (Type-II), and
  random (Type-III)}. Intuitively, a continuous transition occurs when the
system transitions to a nearby local attractor, a catastrophic transition occurs
when there are no nearby local attractors after transition and finally random
transition occurs when the system either moves to a local attractor or leave the
local neighborhood depending on the initial perturbation. For some of the recent
applications of this theory, we refer to \cite{wang2020a, wang2020, lu2019,
  kieu2019, han2019}.

Our analysis shows that four different transitions are possible depending on the
signs of $a_1 b_1$ and $b_3$. When $a_1 b_1 > 0$ there are always bifurcated
saddle mixed mode steady states near the basic solution on both sides of
$\lambda = \lambda_c$. Moreover, the transition is either catastrophic or random
depending on the sign of $b_3$. In the catastrophic transition, there are no
steady states bifurcated from the basic solution after the transition $\lambda >
\lambda_c$ and a repeller bifurcates on $\lambda < \lambda_c$. In the random
transition scenario, the evolution of the system depends on the fluctuations
(initial conditions) of the basic solution. Namely, the phase space separates
into two sectorial regions where solutions starting from the first region leave
the neighbourhood of the basic solution and solutions starting from the second
region tend to an attractor nearby which consists of three steady states and the
orbits between them.

When $a_1 b_1 < 0$, the only bifurcated steady states are the two roll-type
solutions which are symmetric of each other. Depending on the sign of $b_3$,
there is either a continuous transition or a catastrophic transition. In the
continuous transition case, the two roll type solutions are bifurcated on
$\lambda > \lambda_c$ with one being stable and the other being saddle. In this
case, there is an $S^1$ attractor which consists of these two steady states and
the heteroclinic orbits connecting them. In the catastrophic transition case,
there are no steady states bifurcated on $\lambda > \lambda_c$ and two roll type
solutions are bifurcated on $\lambda < \lambda_c$ which form a repeller
homeomorphic to $S^1$.

After the presentation of the general theory, we give three applications which
show that all the transitions described by our main theorems are indeed
observable. We consider two variants of the 2D Swift Hohenberg equation with
quadratic-cubic nonlinearity. The last application is the 3D surface tension
driven convection also known as Marangoni convection. For both systems, we
demonstrate that our main assumptions are satisfied and describe the transition
behavior from the main results we have proved.

\subsection{Notations}
We denote the adjoint eigenvectors by $f_j^*$ which are found by
\begin{displaymath}
  L^* f_j^* = \overline{\beta_j} f_j^*,
\end{displaymath}
where $L^*$ is the adjoint linear operator. The eigenvectors of the linear and
adjoint operator satisfy the orthogonality property
\begin{displaymath}
  \langle f_j, f_k^* \rangle = \delta_{j k} \langle {f_j, f_j^*} \rangle.
\end{displaymath}
with $\langle \cdot, \cdot \rangle$ denoting the inner product in $X$.

Recalling $G_2$ and $G_3$ from \eqref{Taylor}, for ease of notation, we
will denote the bilinear and trilinear interactions of modes by
\begin{equation}\label{non-not}
\begin{aligned}
  & G_2(i,j,k)=\frac{1}{\langle f_k,f_k^{\ast}\rangle}\langle
  G_2(f_i,f_j),f_k^{\ast}\rangle \\
  & G_2^s(i,j,k)=G_2(i,j,k)+G_2(i,k,j)
\end{aligned}
\end{equation}

Similarly for the trilinear operator,
\begin{equation}\label{non-not3}
\begin{aligned}
  & G_3(i,j,k,l)=\frac{1}{\langle f_l,f_l^{\ast}\rangle}\langle
  G_3(f_i,f_j, f_k),f_l^{\ast}\rangle \\
  & G_3^s(i,j,k,l)=\sum_{\sigma}^{} G_3(\sigma(i,j,k),l)
\end{aligned}
\end{equation}
where the summation is over all permutations $\sigma$ of the triple $(i,j,k)$.

\section{Main transition theorems}
In this section, we present the main results on transitions of the system
\eqref{main} under the given conditions. According to dynamic transition theory,
in any dissipative system, there are only three possible transition types. The
intuitive understanding of three types of transitions is mentioned in the
Introduction. For the exact definitions of three types of transitions we refer
to \cite{ptd}.

\subsection{The reduced equations}
We first give the reduced equations with the exact expressions for their
coefficients and mention several remarks regarding these equations.

We denote the center part of the solution spanned by the first two critical
modes by
\[
  u_c = u_1(t) f_1 + u_2(t) f_2
\]
where $f_1$ and $f_2$ given by \eqref{eigenmode-assumption} are the first
critical modes satisfying the PES conditions \eqref{PES} and $u_1(t), u_2(t) \in
\mathbb{R}$ are the time dependent amplitudes of these modes.

In \autoref{appendix}, we prove that the dynamics of near the system near the
origin and close to onset of transition $\lambda = \lambda_c$ is given by the
following reduced equations become
\begin{equation}\label{reduced}
\begin{aligned}
  &\frac{du_1}{dt}=\beta(\lambda) u_1+F_1(x)+O(4) \\
  &\frac{du_2}{dt}=\beta(\lambda) u_2+F_2(x)+O(4)
\end{aligned}
\end{equation}
where the vector field $F=(F_1,F_2)$ is defined as follows:
\begin{equation} \label{F1F2}
\begin{aligned}
& F_1(u_1, u_2)=a_1 u_1 u_2 +u_1 (a_2 u_1^2+a_3 u_2^2), \\
& F_2(u_1, u_2)=b_1 u_1^2 +u_2(b_2 u_1^2+b_3 u_2^2)
\end{aligned}
\end{equation}
and
\[
  O(n)=O(|x|^n)+O(|x|^{n-1}\beta(\lambda)), \quad \text{as } x\rightarrow 0, \lambda\rightarrow\lambda_c.
\]
Here the coefficients of the quadratic terms are
\begin{equation} \label{a1-b1}
    \begin{aligned}
    & a_1 = G_2^s(1,2,1) \\
    & b_1 = G_2(1, 1, 2)
  \end{aligned}
\end{equation}
and the coefficients of the cubic terms are
\begin{equation} \label{a2-b3}
  \begin{aligned}
    a_2 = & G_3(1, 1, 1, 1) + \sum_{j\ge3} \frac{-1}{\beta_j} G_2(1,1,j) G_2^s(1,j,1) \\
    a_3 = & G_3^s(1,2,2,1) +
    \sum_{j\ge3} \frac{-1}{\beta_j} \bigg( G_2(1,2,j) G_2^s(2,j,1) \\
    & + G_2(2,1,j) G_2^s(2,j,1) + G_2(2,2,j) G_2^s(1,j,1) \bigg) \\
    b_2 = & G_3^s(1,1,2,2) +
    \sum_{j\ge3} \frac{-1}{\beta_j} \bigg( G_2(1,1,j) G_2^s(2,j,2) \\
    & + G_2(1,2,j) G_2^s(1,j,2) + G_2(2,1,j) G_2^s(1,j,2) \bigg) \\
    b_3 = & G_3(2,2,2,2) + \sum_{j\ge3} \frac{-1}{\beta_j} G_2(2,2,j) G_2^s(2,j,2)
  \end{aligned}
\end{equation}

\textbf{Remarks}.
\begin{enumerate}
\item We note that the cubic terms depend on the interactions of the first two
  modes with higher modes ($f_j$ with $j\ge3$), while the quadratic terms are
  determined solely by the self-interactions of the first two modes.

\item The reduced equations \eqref{reduced} are symmetric under $u_1 \mapsto -
  u_1$. As a result, the reduced phase portrait is symmetric with respect to
  $u_2$ axis.
\item The first equation of the reduced equation can be solved as
  \begin{displaymath}
    u_1(t) = u_1(0) \exp \left( \int_{0}^{t} \left( a_1 u_2(s) + a_2 u_1^2(s) +
        a_3 u_2^2(s) \right) d s \right)
  \end{displaymath}
  As a result, in the reduced phase portrait, the sign of $u_1(0)$ is preserved
  by $u_1(t)$.
\item When the first two modes are both rolls or both rectangles with equal wave
  numbers, our analysis can be extended to show that quadratic part of the
  reduced equations vanish while the cubic part remains the same. However, our
  main theorems only deal with the case of non-vanishing quadratic terms $a_1$
  and $b_1$. We plan to address the case $a_1 = b_1 = 0$ later.
\item In this paper, we will always assume that $b_3\neq 0$. Otherwise, $( 0, 0
  )$ can easily seen to be non-isolated singular point of $F$.
\item When the critical modes have equal horizontal wave numbers as discussed in
  \ref{sec: setting}, we have the relation $k_1 > j_1$ between the wave indices
  in \eqref{eigenmode-assumption}. In that case, $ C_{11} \cap C_{22} =
  \emptyset $ and the $G_2(2,2,j) G_2^s(1,j,1)$ term in $a_3$ and $G_2(1,1,j)
  G_2^s(2,j,2)$ in $b_2$ also vanish as a result.

  Moreover, when the wave numbers of the critical modes are equal, in the sums
  \eqref{a2-b3}, only modes $f_j$ spanned by $e_{i k}$ with wave indices $(i,
  k)$ in $C_{11}$ for $a_2$, $C_{12}$ for $a_3$, $b_2$ and $C_{22}$ for $b_3$
  have to be considered where
  \begin{equation} \label{Cij}
    \begin{aligned}
      & C_{11} = \left\{ (2j_1, 2j_2), (2j_1, 0), (0, 2j_2) \right\}, \\
      & C_{12} = \{ (k_1-j_1, j_2), (k_1+j_1, j_2) \}, \\
      & C_{22} = \left\{ (2k_1, 0) \right\}.
    \end{aligned}
  \end{equation}

\item The quadratic part of the reduced equations is degenerate and as a result,
  to study the transition behaviour, the cubic part (only $b_3$ term, as we
  shall see) is also necessary.
 
\end{enumerate}

\subsection{The statement of the main theorems}
Before we state the main theorems, we briefly discuss the possible bifurcated
steady state solutions of the system \eqref{main} and give the conditions which
dictate their existence and stability. 
\begin{equation}\label{steady-roll}
\begin{aligned}
  R_1 & = -\sqrt{-\frac{\beta}{b_3}} f_2 + o(\sqrt{-\beta}), \\
  R_2 & = \sqrt{-\frac{\beta}{b_3}} f_2 + o(\sqrt{-\beta})
\end{aligned}
\end{equation}
and
\begin{equation}\label{steady-hex}
\begin{aligned}
  & H_1 =  \frac{1}{\sqrt{a_1 b_1}} \beta f_1 -\frac{1}{a_1} \beta f_2 + o(\beta), \\
  & H_2 =  -\frac{1}{\sqrt{a_1 b_1}} \beta f_1 -\frac{1}{a_1} \beta f_2 + o(\beta),
\end{aligned}
\end{equation}
where $f_1$, $f_2$ are the first critical eigenmodes with eigenvalue $\beta$
satisfying the PES conditions given by \eqref{PES}. First, we
note the scale difference between the steady states $R_i = O(-\sqrt{ \beta })$
and $H_i = O(\beta)$ as $\beta \to 0$. Thus the steady states $H_i$ are much
closer to the basic steady state than $R_i$. Second $R_i$ denote the bifurcated
solutions which have the spatial roll pattern $e_{k_1, 0}$ plus small
perturbations when the eigenmodes are given by \eqref{eigenmode-assumption}. These
solutions are defined only when $\beta(\lambda)b_3<0$. Third, $H_i$ denote the
mixed mode bifurcated steady state solutions which may spatially represent
hexagonal patterns. These solutions are defined when $a_1b_1>0$.

As is well known the transitions of the system are captured by the reduced
equations \eqref{reduced} which contains 6 coefficients determined by the
system. In the case $a_1 \ne 0$, $b_1 \ne 0$, the critical modes have self
interactions. In addition, if $b_3 \ne 0$, we will show that the transition of
the system depends only on the signs of the 3 out of 6 coefficients $a_1$, $b_1$
and $b_3$ in \eqref{reduced}. In this case, the remaining coefficients $a_2$,
$a_3$ and $b_2$ play a quantitative but not a qualitative role. Our main
theorems classify the transitions depending on the signs of these three
coefficients as well as the stability of the above steady states. We summarize
this in \autoref{tabtrans} and \autoref{tabtrans2}.

\begin{table}[ht]
  \begin{tabular}{|c|c|c|c|c|}
    \hline
    $a_1 b_1$ & $b_3$ & Theorem & Transition Type  & Transition Diagram \\ \hline
    $>0$ &$<0$ & \autoref{theorema1b1>0}(i)  & random & \autoref{figa1b1>0b3<0} \\
    $>0$ &$>0$ & \autoref{theorema1b1>0}(ii) & catastrophic & \autoref{figa1b1>0b3>0} \\
    $<0$ &$<0$ & \autoref{theorema1b1<0}(i)  & continuous & \autoref{figa1b1<0b3<0} \\
    $<0$ &$>0$ & \autoref{theorema1b1>0}(ii) & catastrophic & \autoref{figa1b1<0b3>0} \\
    \hline
  \end{tabular}
  \caption{The type of transition and the conditions required. \label{tabtrans}}
\end{table}

\begin{table}[ht]
\begin{tabular}{|c|c|p{2.5cm}|p{3cm}|p{3cm}|}
\hline
 $a_1 b_1$ & $b_3$ & $H_i$ on $\lambda \ne \lambda_c$
  & $R_i$ on $\lambda < \lambda_c$ & $R_i$ on $\lambda > \lambda_c$ \\ \hline
 $>0$ &$<0$ & both are SAD & DNE & $R_1$ is a SN, \qquad $R_2$ is a SAD \\ \hline
 $>0$ &$>0$ & both are SAD & $R_1$ is a SAD, \,\,\, $R_2$ is an UN & DNE \\ \hline
 $<0$ &$<0$ & DNE & DNE & $R_1$ is a SN, \qquad $R_2$ is a SAD \\ \hline
 $<0$ &$>0$ & DNE & $R_1$ is a SAD, \,\, $R_2$ is an UN & DNE \\
 \hline
\end{tabular}
\caption{Stability properties of the bifurcated steady states for $a_1 > 0$. For
  $a_1 < 0$, the stability of $H_i$ is unchanged while the stability of $R_1$
  and $R_2$ is exchanged. Here SAD = saddle, SN = stable node, UN = unstable
  node, DNE = does not exist.\label{tabtrans2}}
\end{table}

Under the conditions given in \autoref{sec: setting}, the transition behavior of
the general dissipative system \eqref{main} is given by the following two
theorems.
\begin{theorem} \label{theorema1b1>0}
 Assume $a_1 b_1>0$.
\begin{itemize}
\item[i)] If $b_3<0$ then the system \eqref{main} undergoes a random (Type-III)
  transition at $\lambda=\lambda_c$ described as below.
\begin{itemize}
\item[a)] The topological structure of the transition is as in \autoref{b3<0}.

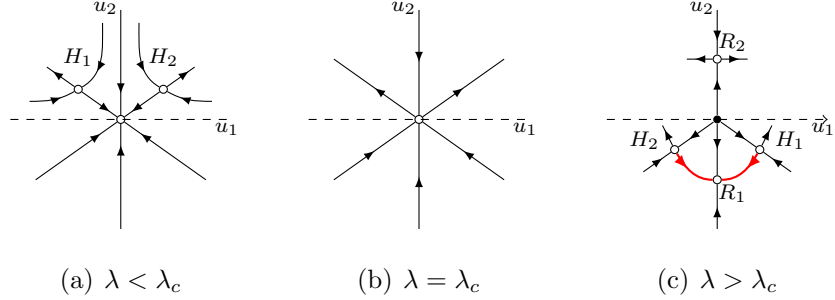
\begin{figure}[H]
\centering
\subfigure[$\lambda<\lambda_c$]{
\begin{tikzpicture}[scale=.8]
  \tikzstyle{every node}=[font=\tiny]
  \node[circle,fill=white, draw=black, ,inner sep=0pt,minimum size=3pt] (O) at (0,0) {};
  \node[circle,fill=white, draw=black,inner sep=0pt,minimum size=3pt,label={[xshift=0mm, yshift=1mm]$H_1$}] (H1) at (-0.7,0.5) {};
  \node[circle,fill=white, draw=black,inner sep=0pt,minimum size=3pt,label={[xshift=0mm, yshift=1mm]$H_2$}] (H2) at (0.7,0.5) {};
  \node[] (2H1) at (-1.4,1) {};
  \node[] (2H2) at (1.4,1) {};
  \node[label={[xshift=-2mm, yshift=-5mm]$u_1$}] (u1) at (2, 0) {};
  \node[label={[xshift=-2mm, yshift=-5mm]$u_2$}] (u2) at (0, 2) {};
  \node[] (nu1) at (-2, 0) {};
  \node[] (nu2) at (0, -2) {};
  \draw[dashed] (nu1) to (u1);
  \draw[->-=.8] (u2) to (O);
  \draw[->-=.8] (nu2) to (O) ;
  \draw[->-=.8] (H1) to (O);
  \draw[->-=.8] (H2) to (O);
  \draw[->-=.8] (H1) to (2H1);
  \draw[->-=.8] (H2) to (2H2);
  \draw[->-=.8] (-1.4,-1) to (O);
  \draw[->-=.8] (1.4,-1) to (O);
  \draw[->-=.7] (-1.5, .3) to [in=210,out=0]  (H1) ;
  \draw[->-=.7] (-.3, 1.6) to [in=30,out=270]  (H1) ;
  \draw[->-=.7] (1.5, .3) to [in=330,out=180] (H2) ;
  \draw[->-=.7] (.3, 1.6) to [in=150,out=270] (H2) ;
 \end{tikzpicture}
}
\subfigure[$\lambda=\lambda_c$]{
\begin{tikzpicture}[scale=.8]
  \tikzstyle{every node}=[font=\tiny]
  \node[circle,fill=white, draw=black,inner sep=0pt,minimum size=3pt] (O) at (0,0) {};
  \node[label={[xshift=-2mm, yshift=-5mm]$u_1$}] (u1) at (2, 0) {};
  \node[label={[xshift=-2mm, yshift=-5mm]$u_2$}] (u2) at (0, 2) {};
  \node[] (nu1) at (-2, 0) {};
  \node[] (nu2) at (0, -2) {};
  \draw[dashed] (nu1) to (u1);
  \draw[->-=.5] (u2) to (O) ;
  \draw[->-=.5] (nu2) to (O);
  \draw[->-=.5] (O) to (-1.4,1);
  \draw[->-=.5] (O) to (1.4,1);
  \draw[->-=.5] (-1.4,-1) to (O);
  \draw[->-=.5] (1.4,-1) to (O);
  \end{tikzpicture}

}
\subfigure[$\lambda>\lambda_c$]{
  \begin{tikzpicture}[scale=.8]
    \tikzstyle{every node}=[font=\tiny]
    \node[circle,fill=black,inner sep=0pt,minimum size=3pt] (O) at (0,0) {};
    \node[circle,fill=white,draw=black,inner sep=0pt,minimum size=3pt,
    label={[xshift=4mm, yshift=-2mm]$H_1$}] (H1) at (0.7,-0.5) {};
    \node[circle,fill=white,draw=black,inner sep=0pt,minimum size=3pt,
    label={[xshift=-4mm, yshift=-2mm]$H_2$}] (H2) at (-0.7,-0.5) {};
    \node[circle,draw=black,fill=white,inner sep=0pt,minimum size=3pt,
    label={[xshift=2mm, yshift=-5mm]$R_1$}] (R1) at (0,-1) {};
    \node[circle,draw=black,fill=white,inner sep=0pt,minimum size=3pt,
    label={[xshift=2mm, yshift=-1mm]$R_2$}] (R2) at (0,1) {};
    \node[] (2H1) at (1.4,-1) {};
    \node[] (2H2) at (-1.4,-1) {};
    \node[label={[xshift=-2mm, yshift=-5mm]$u_1$}] (u1) at (2, 0) {};
    \node[label={[xshift=-2mm, yshift=-5mm]$u_2$}] (u2) at (0, 2) {};
    \node[] (nu1) at (-2, 0) {};
    \node[] (nu2) at (0, -2) {};
    \draw[dashed, ->=1] (nu1) to (u1);
    \draw[->-=.7] (u2) to (R2) ;
    \draw[->-=.7] (O) to (R2) ;
    \draw[->-=.5] (O) to (R1) ;
    \draw[->-=.5] (nu2) to (R1) ;
    \draw[->-=.7] (O) to (H2);
    \draw[->-=.7] (O) to (H1);
    \draw[->-=.7] (2H1) to (H1);
    \draw[->-=.7] (2H2) to (H2);
    \draw[->-=.8] (H2) to (-.9,-.1) ;
    \draw[->-=.8] (H1) to (.9,-.1) ;
    \draw[->-=.4, red, thick] (H2) to  [in=180,out=-60] (R1)    ;
    \draw[->-=.4, red, thick] (H1) to [in=0,out=240]  (R1)     ;
    \draw[->-=.7,] (R2) to  [in=180,out=0] (0.5,1)    ;
    \draw[->-=.7,] (R2) to  [in=0,out=180] (-0.5,1)    ;
  \end{tikzpicture}
\label{attract}
}
\caption{The structure of the transition for $a_1 b_1 > 0$, $b_3<0, a_1>0$. The
  bifurcated attractor $\Sigma_{\lambda}$ on $\lambda>\lambda_c$ is shown in
  red. When $a_1<0$, the assertions given by \autoref{theorema1b1>0} hold true
  with the regions and the steady states flipped with respect to the $u_1$
  axis. \label{figa1b1>0b3<0}}
\label{b3<0}
\end{figure}

\item[b)] There is a neighborhood $\mathcal{U}$ of $\phi=0$ in the phase space
  $X$ such that for any $\lambda_c<\lambda<\lambda_c+\epsilon$ with some
  $\epsilon>0$, $\mathcal{U}$ can be decomposed into two open sets
  $\mathcal{U}_1^{\lambda}$, $\mathcal{U}_2^{\lambda}$,
  \begin{displaymath}
    \overline{\mathcal{U}}=\overline{\mathcal{U}_1^{\lambda}}\cup\overline{\mathcal{U}_2^{\lambda}},
    \qquad \mathcal{U}_1^{\lambda}\cap\mathcal{U}_2^{\lambda}=\emptyset
  \end{displaymath}
  such that
  \begin{equation*}
    \begin{aligned}
      & \lim_{\lambda\rightarrow\lambda_c}\limsup_{t\rightarrow\infty} \norm{ S_{\lambda}(t,\varphi) }_X=0
      && \qquad \forall \varphi\in\mathcal{U}_1^{\lambda}, \\
      & \limsup_{t\rightarrow\infty} \norm{ S_{\lambda}(t,\varphi) }_X\geq\delta>0
      && \qquad \forall \varphi\in\mathcal{U}_2^{\lambda},
    \end{aligned}
  \end{equation*}
  for some $\delta>0$. Here $S_{\lambda}$ is the evolution of the solution with
  initial data $\varphi$. Moreover $\mathcal{P}(U_1^\lambda)$,
  $\mathcal{P}(U_2^\lambda)$ are sectorial regions as shown in \autoref{attract}
  with angles $\pi - 2\theta$ and $\pi + 2\theta$ respectively, where
  $\theta=\arctan{\sqrt{a_1/b_1}}$ and $\mathcal{P}$ is the projection onto the
  plane spanned by $f_1,f_2$.
\item[c)] The system bifurcates to an attractor $\Sigma_{\lambda}$ which
  consists of three steady states $R_1,H_1,H_2$ and the heteroclinic orbits
  connecting $H_i$ to $R_1,i=1,2$. Namely, $\Sigma_{\lambda}$ is the arc
  connecting these three steady states as shown in \autoref{attract}, and has
  $\mathcal{U}_1^{\lambda}$ as its basin of attraction.
\end{itemize}
\item[ii)] If $b_3>0$ then the system undergoes a catastrophic (Type-II) transition at
  $\lambda=\lambda_c$ and the following assertions are true:
\begin{itemize}
\item[a)] The topological structure of the transition is as given by
  \autoref{figa1b1>0b3>0}.

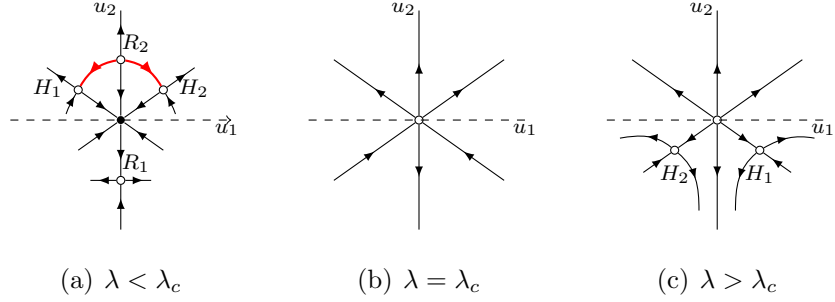
\begin{figure}[H]
\centering
\subfigure[$\lambda<\lambda_c$]{
\begin{tikzpicture}[scale=.8]
  \tikzstyle{every node}=[font=\tiny]
  \node[circle,fill=black,inner sep=0pt,minimum size=3pt] (O) at (0,0) {};
  \node[circle,fill=white,draw=black,inner sep=0pt,minimum size=3pt,
  label={[xshift=-4mm, yshift=-3mm]$H_1$}] (H1) at (-0.7,0.5) {};
  \node[circle,fill=white,draw=black,inner sep=0pt,minimum size=3pt,
  label={[xshift=4mm, yshift=-3mm]$H_2$}] (H2) at (0.7,0.5) {};
  \node[circle,draw=black,fill=white,inner sep=0pt,minimum size=3pt,
  label={[xshift=2mm, yshift=-1mm]$R_1$}] (R1) at (0,-1) {};
  \node[circle,draw=black,fill=white,inner sep=0pt,minimum size=3pt,
  label={[xshift=2mm, yshift=-1mm]$R_2$}] (R2) at (0,1) {};
  \node[] (2H2) at (1.4,1) {};
  \node[] (2H1) at (-1.4,1) {};
  \node[label={[xshift=-2mm, yshift=-5mm]$u_1$}] (u1) at (2, 0) {};
  \node[label={[xshift=-2mm, yshift=-5mm]$u_2$}] (u2) at (0, 2) {};
  \node[] (nu1) at (-2, 0) {};
  \node[] (nu2) at (0, -2) {};
  \draw[dashed, ->=1] (nu1) to (u1);
  \draw[->-=.7] (R2) to (u2);
  \draw[->-=.7] (R2) to (O);
  \draw[->-=.7] (O) to (R1);
  \draw[->-=.7] (nu2) to (R1);
  \draw[->-=.7] (H2) to (O);
  \draw[->-=.7] (H1) to (O);
  \draw[->-=.7] (H1) to (2H1);
  \draw[->-=.7] (H2) to (2H2);
  \draw[->-=.7] (-.7, -.5) to (O);
  \draw[->-=.7] (.7, -.5) to (O);
  
  \draw[->-=.7, red, thick] (R2) to  [in=120,out=-10] (H2)     ;
  \draw[->-=.7, red, thick] (R2) to [in=60,out=190]  (H1)     ;
  \draw[->-=.9] (.9,.1) to [in=-60,out=110] (H2)  ;
  \draw[->-=.9]  (-.9,.1) to [in=240,out=70] (H1) ;
  \draw[->-=.7,] (R1) to  [in=180,out=0] (0.5,-1)    ;
  \draw[->-=.7,] (R1) to  [in=0,out=180] (-0.5,-1)    ;
\end{tikzpicture}
}
\subfigure[$\lambda=\lambda_c$]{
\begin{tikzpicture}[scale=.8]
  \tikzstyle{every node}=[font=\tiny]
  \node[circle,fill=white, draw=black,inner sep=0pt,minimum size=3pt] (O) at (0,0) {};
  \node[label={[xshift=-2mm, yshift=-5mm]$u_1$}] (u1) at (2, 0) {};
  \node[label={[xshift=-2mm, yshift=-5mm]$u_2$}] (u2) at (0, 2) {};
  \node[] (nu1) at (-2, 0) {};
  \node[] (nu2) at (0, -2) {};
  \draw[dashed] (nu1) to (u1);
  \draw[->-=.5] (O) to (u2);
  \draw[->-=.5] (O) to (nu2);
  \draw[->-=.5] (O) to (-1.4,1);
  \draw[->-=.5] (O) to (1.4,1);
  \draw[->-=.5] (-1.4,-1) to (O);
  \draw[->-=.5] (1.4,-1) to (O);
  \end{tikzpicture}
}
\subfigure[$\lambda>\lambda_c$] {
\begin{tikzpicture}[scale=.8]
  \tikzstyle{every node}=[font=\tiny]
  \node[circle,fill=white, draw=black, inner sep=0pt,minimum size=3pt] (O) at (0,0) {};
  \node[circle,fill=white, draw=black, inner sep=0pt,minimum size=3pt,
  label=below:{$H_1$}] (H1) at (0.7,-0.5) {};
  \node[circle,fill=white, draw=black,inner sep=0pt,minimum size=3pt,
  label=below:{$H_2$}] (H2) at (-0.7,-0.5) {};
  \node[] (2H1) at (1.4,-1) {};
  \node[] (2H2) at (-1.4,-1) {};
  \node[label={[xshift=-2mm, yshift=-5mm]$u_1$}] (u1) at (2, 0) {};
  \node[label={[xshift=-2mm, yshift=-5mm]$u_2$}] (u2) at (0, 2) {};
  \node[] (nu1) at (-2, 0) {};
  \node[] (nu2) at (0, -2) {};
  \draw[dashed] (nu1) to (u1);
  \draw[->-=.5] (O) to (u2);
  \draw[->-=.5] (O) to (nu2);
  \draw[->-=.8] (O) to (H1);
  \draw[->-=.8] (O) to (H2);
  \draw[->-=.7] (2H1) to (H1);
  \draw[->-=.7] (2H2) to (H2);
  \draw[->-=.5] (O) to (1.4,1);
  \draw[->-=.5] (O) to (-1.4,1);
  \draw[->-=.5] (H1) to [in=90,out=220] (.3, -1.5)  ;
  \draw[->-=.5] (H1) to [in=170,out=30] (1.6, -.3)  ;
  \draw[->-=.5] (H2) to [in=90,out=-40] (-.3, -1.5) ;
  \draw[->-=.5] (H2) to [in=10,out=140] (-1.6, -.3) ;
 \end{tikzpicture}
}
\caption{The structure of transition for $a_1 b_1 > 0$, $b_3>0, a_1>0$. The
  bifurcated repeller $\Sigma_{\lambda}$ on $\lambda<\lambda_c$ is shown in red
  color. For $a_1 < 0$, the same transition diagram is obtained with $H_1$,
  $H_2$ solutions reflected along $u_1$ axis for $\lambda>
  \lambda_c$. \label{figa1b1>0b3>0}}
\end{figure}

\item[b)] There is a bifurcated repeller
  $\Sigma_{\lambda}$ on $\lambda<\lambda_c$ which consists of three steady
  states, $H_1, H_2, R_2$ and the heteroclinic orbits connecting $R_2$ to $H_1$
  and $H_2$ respectively. $\Sigma_{\lambda}$, topologically.
\item[c)] Finally for $\lambda_c+\epsilon>\lambda>\lambda_c$ there is an open
  neighborhood $\mathcal{U}$ of $\phi=0$ and a dense, open subset
  $\mathcal{U}^{\lambda}$ of $\mathcal{U}$ such that
$$ \limsup_{t\rightarrow\infty} ||S_{\lambda}(t,\varphi)||_X\geq\delta>0, \qquad \forall \varphi\in\mathcal{U}^{\lambda},
$$
for some $\delta>0$.
\end{itemize}
\end{itemize}
\end{theorem}

\begin{theorem}\label{theorema1b1<0}
Assume $a_1 b_1<0$.
\begin{itemize}
\item[i)] If $b_3<0$, then the system \eqref{main} undergoes a continuous
  (Type-I) transition at $\lambda=\lambda_c$ described as below.
\begin{itemize}
\item[a)] The topological structure is as given by \autoref{figa1b1<0b3<0}.

\begin{figure}[H]
\centering
\subfigure[$\lambda\leq\lambda_c$]
{
\begin{tikzpicture}[scale=.8]
  \tikzstyle{every node}=[font=\tiny]
  \node[circle,fill=black,inner sep=0pt,minimum size=3pt] (O) at (0,0) {};
  \node[label={[xshift=-2mm, yshift=-5mm]$u_1$}] (u1) at (2, 0) {};
  \node[label={[xshift=-2mm, yshift=-5mm]$u_2$}] (u2) at (0, 2) {};
  \node[] (nu1) at (-2, 0) {};
  \node[] (nu2) at (0, -2) {};
  \draw[dashed, ->=1] (nu1) to (u1);
  \draw[->-=.7] (u2) to (O);
  \draw[->-=.7] (nu2) to (O);
  \begin{scope}
    \draw[] plot [smooth,tension=1]
    coordinates { (-.3,-2) (-.5, -1) (-.8,0) (-1.2,1) (-.8,1.4)(-.2,.8) (0,0)}
    [arrow inside={end=latex,opt={black,scale=1}}{0.25,0.5,0.75}];
  \end{scope}
  \begin{scope}
    \draw[] plot [smooth,tension=1]
    coordinates { (.3,-2) (.5, -1) (.8,0) (1.2,1) (.8,1.4) (.2, .8)(0,0)}
    [arrow inside={end=latex,opt={black,scale=1}}{0.25,0.5,0.75}];
  \end{scope}
\end{tikzpicture}
}
\subfigure[$\lambda>\lambda_c$]{
\begin{tikzpicture}[scale=.8]
  \tikzstyle{every node}=[font=\tiny]
  \node[circle,fill=black,inner sep=0pt,minimum size=3pt] (O) at (0,0) {};
  \node[circle,draw=black,fill=white,inner sep=0pt,minimum size=3pt,
  label={[xshift=2mm, yshift=-1mm]$R_1$}] (R1) at (0,-1) {};
  \node[circle,draw=black,fill=white,inner sep=0pt,minimum size=3pt,
  label={[xshift=2mm, yshift=-1mm]$R_2$}] (R2) at (0,1) {};
  \node[label={[xshift=-2mm, yshift=-5mm]$u_1$}] (u1) at (2, 0) {};
  \node[label={[xshift=-2mm, yshift=-5mm]$u_2$}] (u2) at (0, 2) {};
  \node[] (nu1) at (-2, 0) {};
  \node[] (nu2) at (0, -2) {};
  \draw[dashed, ->=1] (nu1) to (u1);
  \draw[->-=.7] (u2) to (R2);
  \draw[->-=.7] (O) to (R2);
  \draw[->-=.7] (O) to (R1);
  \draw[->-=.7] (nu2) to (R1);
  
  \draw[->-=.5, red, thick] (R1) to [bend left=90, looseness=1.25] (R2)     ;
  \draw[->-=.5, red, thick] (R1) to [bend right=90, looseness=1.25]  (R2)     ;
\end{tikzpicture}

}
\caption[Optional caption for list of figures]{The structure of transition for $a_1b_1<0,b_3<0$.}
\label{figa1b1<0b3<0}
\end{figure}
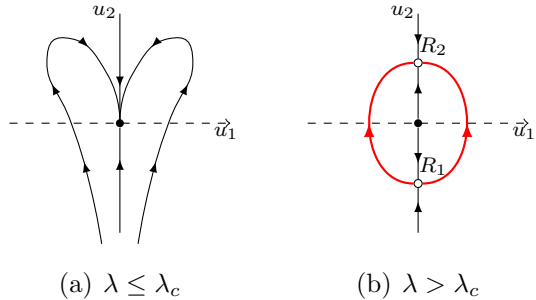

\item[b)] The system bifurcates on $\lambda>\lambda_c$ to an attractor
  $\Sigma_{\lambda}$ which is homeomorphic to $S^1$. $\Sigma_{\lambda}$ consists
  of two singular points $R_i$, $i=1,2$ and two heteroclinic orbits connecting
  them. Moreover there exists a neighborhood $\mathcal{U}^{\lambda}$ of $\phi=0$
  such that $\Sigma_{\lambda}$ attracts $\mathcal{U}^{\lambda}\setminus \Gamma$
  where $\Gamma$ is the stable manifold of $\phi=0$ with $codim\Gamma=2$.
\end{itemize}
\item[ii)] If $b_3>0$, then the system undergoes a catastrophic (Type-I)
  transition at $\lambda=\lambda_c$.
\begin{itemize}
\item[a)] The topological structure is as given by \autoref{figa1b1<0b3>0}.
  \begin{figure}[ht]
\centering
\subfigure[$\lambda<\lambda_c$]{
\begin{tikzpicture}[scale=.8]
  \tikzstyle{every node}=[font=\tiny]
  \node[circle,fill=black,inner sep=0pt,minimum size=3pt] (O) at (0,0) {};
  \node[circle,draw=black,fill=white,inner sep=0pt,minimum size=3pt,
  label={[xshift=2mm, yshift=-1mm]$R_1$}] (R1) at (0,-1) {};
  \node[circle,draw=black,fill=white,inner sep=0pt,minimum size=3pt,
  label={[xshift=2mm, yshift=-1mm]$R_2$}] (R2) at (0,1) {};
  \node[label={[xshift=-2mm, yshift=-5mm]$u_1$}] (u1) at (2, 0) {};
  \node[label={[xshift=-2mm, yshift=-5mm]$u_2$}] (u2) at (0, 2) {};
  \node[] (nu1) at (-2, 0) {};
  \node[] (nu2) at (0, -2) {};
  \draw[dashed, ->=1] (nu1) to (u1);
  \draw[->-=.7] (R2) to (u2);
  \draw[->-=.7] (R2) to (O);
  \draw[->-=.7] (R1) to (nu2);
  \draw[->-=.7] (R1) to (O);
  
  \draw[->-=.5, red, thick] (R2) to [bend left=90, looseness=1.25] (R1)     ;
  \draw[->-=.5, red, thick] (R2) to [bend right=90, looseness=1.25]  (R1)     ;
\end{tikzpicture}
}
\subfigure[$\lambda\geq\lambda_c$]{
\begin{tikzpicture}[scale=.8]
  \tikzstyle{every node}=[font=\tiny]
  \node[circle,fill=black,inner sep=0pt,minimum size=3pt] (O) at (0,0) {};
  \node[label={[xshift=-2mm, yshift=-5mm]$u_1$}] (u1) at (2, 0) {};
  \node[label={[xshift=-2mm, yshift=-5mm]$u_2$}] (u2) at (0, 2) {};
  \node[] (nu1) at (-2, 0) {};
  \node[] (nu2) at (0, -2) {};
  \draw[dashed, ->=1] (nu1) to (u1);
  \draw[->-=.7] (O) to (u2);
  \draw[->-=.7] (O) to (nu2);
  \begin{scope}
    \draw[] plot [smooth,tension=1]
    coordinates {(0,0)(-.2,.8)(-.8,1.4)(-1.2,1)(-.8,0)(-.5, -1)(-.3,-2)}
    [arrow inside={end=latex,opt={black,scale=1}}{0.25,0.5,0.75}];
  \end{scope}
  \begin{scope}
    \draw[] plot [smooth,tension=1]
    coordinates {(0,0)(.2,.8)(.8,1.4)(1.2,1)(.8,0)(.5, -1)(.3,-2)}
    [arrow inside={end=latex,opt={black,scale=1}}{0.25,0.5,0.75}];
  \end{scope}
\end{tikzpicture}
}
\caption{The structure of transition for $a_1b_1<0,b_3>0$.}
\label{figa1b1<0b3>0}
\end{figure}
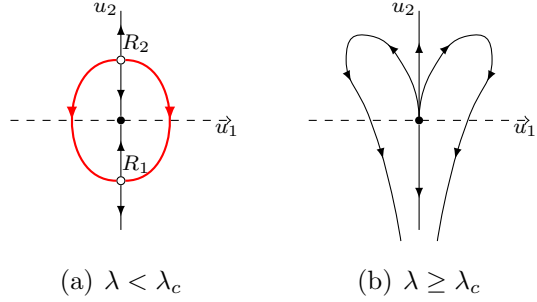

\item[b)] The system bifurcates to a repeller $\Sigma_{\lambda}$ on
  $\lambda<\lambda_c$ which consists of two critical points $R_i$, $i=1,2$ and
  heteroclinic orbits connecting them.
\end{itemize}
\end{itemize}
\end{theorem}

\section{Applications} \label{sec:Applications}
In this section, we give several applications to demonstrate that all the
results obtained in our main theorems are indeed realizable.
\subsection{2D Swift-Hohenberg Equation with quadratic-cubic nonlinearity}
We first consider the 2D Swift-Hohenberg equation
\begin{equation}\label{SHE}
  u_t = \lambda u - ( \Delta + k )^2 u +  \alpha_2 u^2 - \alpha_3 u^3
\end{equation}
with $\alpha_2, \alpha_3, k, \lambda \in \mathbb{R}$ and $u = u(x, y, t)$ is the
unknown function. The Swift-Hohenberg equation (SHE) was first proposed in 1977
\cite{swift1977a} as a simple model for the Rayleigh-Benard instability of roll
waves. For previous results on instabilities and physical aspects of the model,
we refer to \cite{peletier2004, choi2017, li2018, hernandez2018}

We consider a spatial domain
\begin{equation}
  \label{she-dom}
    \Omega = (0, \sqrt{3}\pi) \times (0, \pi),
\end{equation}
and Neumann boundary conditions
\begin{equation} \label{SHE:bc}
\frac{\partial u}{\partial \nu} = \frac{\partial \Delta u}{\partial \nu} =0 \quad
\text{ on }\partial\Omega,
\end{equation}
where $\nu$ is the unit outward normal to $\partial\Omega$.
We also assume zero mean conditions
\begin{equation} \label{SHE:zero-mean}
    \iint_{\Omega} u dx dy = 0.
\end{equation}
For the functional setting, let
\begin{equation} 
    \begin{aligned}
    H&=\{ u\in L^2(\Omega) \ |  \int_{\Omega} u \, \mathrm{d} x = 0 \}, \\
    H_1&=\{u \in H^4(\Omega) \cap H \ |\
    \text{ $u$ satisfies \eqref{SHE:bc} and \eqref{SHE:zero-mean} } 
    \}.
  \end{aligned}
\end{equation}
Let $L_{\lambda}:H_1\rightarrow H$ and $G:H_1\rightarrow H$ be defined by
\begin{equation} \label{operator}
\begin{aligned}
L_{\lambda} u &= \lambda u - (\Delta + k)^2 u, \\
G(u) &= \alpha_2 u^2 - \alpha_3 u^3.
\end{aligned}
\end{equation}
Then the problem (\ref{SHE}) with \eqref{SHE:bc} can be written as
\begin{equation} \label{SHE abstract form}
\begin{aligned}
&\frac{\mathrm{d} u}{\mathrm{d} t}=L_{\lambda} u +G (u),\\
&u(0)=u_0.
\end{aligned}
\end{equation}

Under these conditions, the eigenvalue problem
\[
   \lambda u  - ( \Delta + k )^2u = \beta u
\]
has eigenfunctions
\[
  e_{j_1, j_2}(x, y) = \cos (j_1 \frac{x}{\sqrt{3}}) \cos (j_2 y)
\]
and eigenvalues
\[
  \beta_{j_1, j_2} = \lambda - (k - \abs{ J }^2 )^2, \qquad
  J = (j_1, j_2) \in \mathcal{J}
\]
Here
\begin{displaymath}
  \abs{ (j_1, j_2) }^2 = \frac{ j_1^2 }{ 3 } + j_2^2
\end{displaymath}

\begin{displaymath}
  \mathcal{J} = \{ (j_1, j_2) \in \mathbb{Z}_{\ge 0} \times \mathbb{Z}_{\ge 0} :
  ( j_1, j_2 ) \ne (0, 0) \}
\end{displaymath}

It is easy to see that the PES condition is satisfied with
\[
  \lambda_c(k) = \min_{J \in \mathcal{J}} (k - \abs{ J }^2)^2 =
  (k - \abs{ J_i}^2)^2, \qquad i = 1, \dots, n
\]

The minimum occurs at $\abs{ J }^2 = k$ when $k \in \mathcal{J}$. If $k \notin
\mathcal{J}$, then it occurs at one of $k_1, k_2 \in \mathcal{J}$ for which $k_1
< k < k_2$.
Note that
\begin{displaymath}
  \abs{ (1, 0) } < \abs{ (0, 1) } < \abs{ (1, 1) } = \abs{ (2, 0) } < \abs{ (2,
    1) } < \cdots
\end{displaymath}
Thus solving $(k - \abs{ J }^2 )^2 = (k - \abs{ \tilde{J} }^2)^2$ for $k$ for
the first two consecutive $\abs{ J }$, $\abs{ \tilde{J} }$ gives the following
result.





\begin{displaymath}
  \begin{cases}
    J_1 = (1,0)  & \text{if } 0 < k \le 4/6 \\
    J_1 = (0,1)  & \text{if } 4/6 \le k \le 7/6 \\
    J_1 = (1,1), \, J_2 = (2,0)  & \text{if } 7/6 \le k \le 11/6 \\
    J_1 = (2,1)  & \text{if } 11/6 \le k < 19/6
  \end{cases}
\end{displaymath}
In particular, note that when
\begin{equation} \label{she-conds}
  7/6 < k < 11/6, \quad \lambda_c = (k - 4/3)^2,
\end{equation}
there are two critical modes
\begin{displaymath}
  f_1 = e_{1, 1} = \cos \frac{ x }{ \sqrt{ 3 } } \cos y, \qquad 
  f_2 = e_{2, 0} = \cos \frac{ 2 x }{ \sqrt{ 3 } },
\end{displaymath}
with corresponding eigenvalues satisfying the PES condition 
\begin{equation} \label{SHE:PES}
    \begin{aligned}
    \beta_{1,1} = \beta_{2,0} = \lambda - \lambda_c
    \begin{cases}
    <0, & \lambda<\lambda_c \\
    =0, & \lambda=\lambda_c \\
    >0, & \lambda>\lambda_c
    \end{cases} \\
    \beta_{j_1, j_2} < 0, \qquad (j_1, j_2) \notin \{ (1,1), (2,0) \}.
  \end{aligned}
\end{equation}
Thus the main assumptions are all satisfied.
Hence, the PDE system can be reduced to the ODE system \eqref{reduced} near
$\lambda=\lambda_c$ and for small $(u_1, u_2)$ where $u_1$, $u_2$ are the time
dependent amplitudes of the modes $f_1$, $f_2$.

Now we can compute the coefficients \eqref{a1-b1}, \eqref{a2-b3} that determines
the type of transition. We let
\[
  G_2(i,j,k) = \alpha_2 \int_{\Omega} f_i f_j f_k dx dy
\]
\[
  G_3(i,j,k,l) = \alpha_3 \int_{\Omega} f_i f_j f_k f_l dx dy
\]



\begin{figure}
  \centering
  \includegraphics[scale=.5]{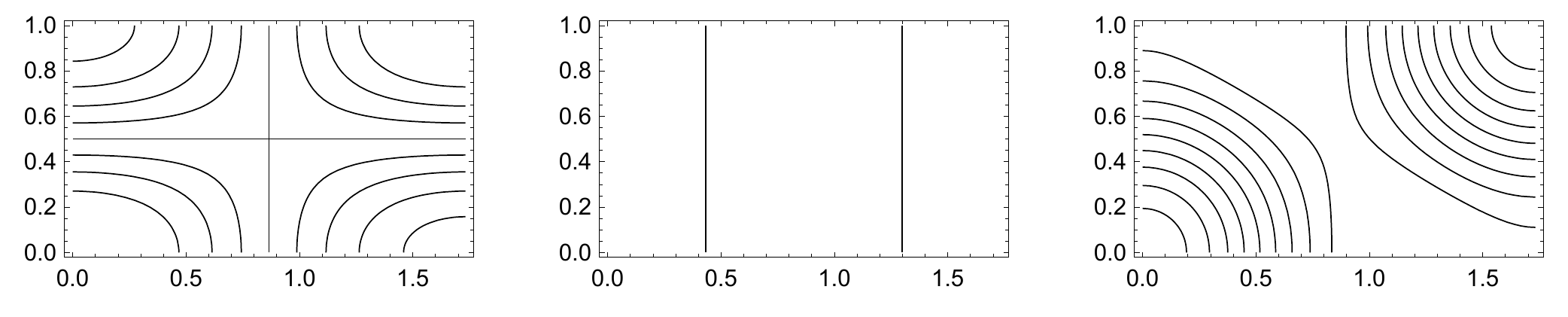}
  \caption{The bifurcated solutions.}
  \label{fig:11rec-rollmode}
\end{figure}

Hence
\[
  a_1 = \alpha_2, \qquad b_1 = \frac{\alpha_2}{4}, \qquad b_3 =
  \frac{3(\alpha_2^2 + 4(3k-10)\alpha_3)}{(160-48k)}.
\]
Note that $a_1 b_1 \ge 0$. Also if $\alpha_2 \ne 0$ then necessarily $a_1 b_1 >
0$. For the range of $k$ values we are interested, $7/6<k<11/6$,
\begin{displaymath}
  b_3 \sim \alpha_2^2 + 4(3k-10) \alpha_3 
\end{displaymath}
and $b_3 > 0$ if $\alpha_3 < 0$ and can be of both signs when $\alpha_3 > 0$.
\begin{theorem}
  Under the conditions \eqref{she-conds} and $\alpha_2 \ne 0$, the basic
  solution $u = 0$ of the equation \eqref{SHE abstract form} undergoes a
  transition at $\lambda = \lambda_c$. The type of transition is either
  continuous if $b_3 < 0$ or catastrophic if $b_3 > 0$ as described by
  \autoref{theorema1b1>0}. In particular, if $\alpha_3 \le 0$ then $b_3 > 0$,
  while for $\alpha_3 > 0$, depending on $\alpha_2$, $b_3$ can be of both signs,
  where $\alpha_2$ and $\alpha_3$ are the coefficients of the bilinear and
  trilinear terms in \eqref{SHE}. The structure of the bifurcated mixed modes
  are as given in \autoref{fig:11rec-rollmode}.
\end{theorem}

\subsection{Modified 2D Swift Hohenberg Equation}
\label{sec:reac-diff}

We will now show that both transition scenarios described by
\autoref{theorema1b1<0} are possible. For this let us consider the following
equation which is more general than \eqref{SHE}.
\begin{equation}
  \label{mSHE main}
  u_t = \lambda u - ( \Delta + k )^2 u + G(u, u) - \alpha_3 u^3
\end{equation}
where the bilinear operator is
\begin{equation}
  \label{mSHE nonlin}
  G(u, v) = (c_1 + c_2 u_x + c_3 u_y) v + u (c_4 v_x + c_5 v_y)
\end{equation}
on the rectangular domain \eqref{she-dom} with the following boundary conditions
\begin{equation}
  \label{mSHE bc}
  \begin{aligned}
    & u = \Delta u = 0, \qquad x=0, \sqrt{ 3 } \pi, \\
    & u_y = \Delta u_y = 0, \qquad y=0, \pi.
  \end{aligned}
\end{equation}
This time, the basis functions are
\[
  e_{j_1, j_2}(x, y) = \sin (j_1 \frac{x}{\sqrt{3}}) \cos (j_2 y)
\]
\begin{equation}
  \label{mSHE conds}
  5/6 < k < 11/6, \qquad \lambda_c = (k - 4/3)^2,
\end{equation}
and there are two critical modes
\begin{displaymath}
  f_1 = e_{1, 1} = \sin \frac{ x }{ \sqrt{ 3 } } \cos y, \qquad 
  f_2 = e_{2, 0} = \sin \frac{ 2 x }{ \sqrt{ 3 } },
\end{displaymath}
with corresponding eigenvalues satisfying the PES condition \eqref{SHE:PES}.
In this case, the numbers describing the transition are found to be
\begin{displaymath}
  a_1 = - \frac{ (c_2 + c_4)^2 }{ 2 \sqrt{ 3 } }, \qquad
  b_1 = \frac{ c_2 + c_4 }{ 4 \sqrt{ 3 } }, \qquad
  b_3 = \frac{ (c_2 + c_4)^2 + 6 (10 -3k) \alpha_3 }{ 24k - 80 }
\end{displaymath}
It is readily seen that for $c_2 + c_4 \ne 0$, $a_1 b_1 < 0$ and the transition
is described by the following theorem.
\begin{theorem}
  Under the conditions \eqref{mSHE conds} and $c_2 + c_4 \ne 0$, the basic
  solution $u = 0$ of the equation \eqref{mSHE main} with \eqref{mSHE bc}
  undergoes a transition at $\lambda = \lambda_c$. The type of transition is
  either continuous if $b_3 < 0$ or catastrophic if $b_3 > 0$ as described by
  \autoref{theorema1b1<0}. In particular, if $\alpha_3 \ge 0$ then $b_3 < 0$,
  while for $\alpha_3 < 0$, depending on $c_2+c_4$, $b_3$ can be of both signs,
  where $c_2$, $c_4$ and $\alpha_3$ are the coefficients of the bilinear and
  trilinear terms in \eqref{mSHE main}.
\end{theorem}

\subsection{Surface tension driven convection}
\label{sec: marangoni convection}

In this section, we present known results on the hexagonal pattern formation in
surface tension driven convection also known as Marangoni convection to show
that it fits into the framework we present in this study. The nondimensional
form of the equations describing the Marangoni convection without gravity are,
\cite{dauby1993, wang2018},
\begin{equation}  \label{Marangoni-main}
\begin{aligned} & \frac{\partial \mathbf{u}}{\partial t}+\left(
\mathbf{u}\cdot \nabla \right) \mathbf{u} = \text{Pr}\left( -\nabla p+\Delta
\mathbf{u}\right) , \\ & \frac{\partial \theta }{\partial t}+\left(
\mathbf{u}\cdot \nabla \right) \theta = w+\Delta \theta , \\ & \nabla \cdot
\mathbf{u} =0. \\ 
\end{aligned}
\end{equation}
For the physical description and results about previous results on the
instabilities of the above system, we refer to \cite{pearson1958, rosenblat1982,
  rosenblat1982a, dauby1993, dauby1996, dijkstra1998, colinet2001,
  dijkstra2013}. Here $\mathbf{u}=\left( u,v,w \right) $ is the velocity field,
$T$ is the temperature, $p$ is the pressure, $\text{Pr}={\nu }/{\kappa }>0$ is
the Prandtl number. The unknowns represent a deviation from the motionless basic
solution with a linear temperature profile given by,
\[
  \begin{aligned}
    & \mathbf{u}_b =0, \\
    & T_b=T_{0}+( T_1-T_0) z, \\
    & p_b(z) = p_0 +g \rho_0 (z-\alpha(T_1 - T_0) z^2/2),
  \end{aligned}
\]  
where $T_0$, $T_1$ are the temperatures at $z = 0$ and $z = 1$ respectively.

We consider the equations \eqref{main} on a rectangular domain
\[
\Omega =\left( 0,L_{1}\right)
\times \left( 0,L_{2}\right) \times \left( 0,1\right) \subset \mathbb{R}^{3}.
\]

We supplement the above system with free-slip boundary conditions on the lateral
boundaries, and the rigid (no slip) boundary condition and perfectly conducting
on the bottom boundary. The top surface is assumed to be a non-deformable free
surface with a surface tension of the form
$$
\xi=\xi_{0} (1 - \gamma_{T}\theta).
$$
Namely, the boundary conditions are as follows:
\begin{equation}  \label{bc}
\begin{aligned} 
& u =\frac{\partial v}{\partial x}=\frac{\partial w}{\partial x}=\frac{\partial \theta }{\partial
x}=0\,
      &&\text{at}\,x=0,L_{1}, \\ & \frac{\partial u}{\partial y}
=v=\frac{\partial w}{\partial y}=\frac{\partial \theta }{\partial
y}=0\,&&\text{at}\,y=0,L_{2}, \\ & u =v=w=\theta =0\,&&\text{at}\,z=0, \\ &
\frac{\partial \left( u,v\right) }{\partial z}+\lambda \nabla _{H}\theta
=w=\frac{\partial \theta }{\partial z}+ \text{Bi} \theta
=0\,&&\text{at}\,z=1, 
\end{aligned}
\end{equation}
where $\nabla _{H}=\left( \partial _{x},\partial _{y}\right) $, $\text{Bi} \geq
0$ is the Biot number, and the Marangoni number $\lambda $ is the control
parameter which represents the ratio of the destabilizing surface tension
gradient to the stabilizing forces associated with thermal and viscous
diffusion.

By the separation of variables, we represent the solutions 
in the following form:
\begin{equation} \label{sepofvar}
\begin{aligned} 
& u_I =U_I\left( z\right) \sin L _{1}^{-1}I_x\pi x\cos L _{2}^{-1}I_y\pi y, \\ 
& v_I =V_I\left( z\right) \cos L _{1}^{-1}I_x\pi x\sin L _{2}^{-1}I_y\pi y, \\ 
& w_I =W_I\left( z\right) \cos L _{1}^{-1}I_x\pi x\cos L _{2}^{-1}I_y\pi y, \\ 
& \theta_I =\Theta_I \left( z\right) \cos L _{1}^{-1}I_x\pi x\cos L _{2}^{-1}I_y\pi y,
\end{aligned}  
\end{equation} 
for $I=(I_x,I_y)\in \mathbb{Z}\times\mathbb{Z}$. If instead of free-slip
boundaries, no-slip boundaries are considered then the corresponding eigenvalue
problem has to be solved by numerical methods \cite{dauby1996, sengul2015}.

In \cite{dijkstra2013}, the following are proved:
\begin{enumerate}
\item When the relation \eqref{L1L2rel} between the horizontal length scales of
  the domain is satisfied, it is possible that two modes with indices $(j_1,
  j_2)$ and $(0, 2j_2)$ are the first two critical modes satisfying the PES
  condition \eqref{PES} at the critical Marangoni number $\lambda_c$, introduced
  in \cite{pearson1958} is defined as
  \begin{displaymath}
    \lambda _{c}=\min_{\alpha} \frac{8\alpha \left( \alpha \cosh \alpha +\text{Bi}\sinh \alpha
      \right) \left( \alpha -\cosh \alpha \sinh \alpha \right) }{\alpha ^{3}\cosh
      \alpha -\sinh ^{3}\alpha },
  \end{displaymath}%
  where the minimum is taken over wave numbers
  \begin{displaymath}
    \alpha = \alpha_{j, k} = \left((L _{1}^{-1} j )^2+(L _{2}^{-1} k )\right)^{1/2}\pi.
  \end{displaymath}

The map of critical index selection is similar to \eqref{fig:wavemap}.

\item Under the above setting, the system has a transition described by the
  reduced equations \eqref{reduced}. Moreover the coefficients of the reduced
  equations satisfy
  \begin{displaymath}
    a_1 = 4b_1, \qquad a_3 = 2b_3, \qquad 4a_2 = a_3 + b_2
  \end{displaymath}
\item Thus $a_1 b_1 > 0$ and the transition is as described by
  \autoref{theorema1b1>0}. The sign of $b_3$ is found to be negative in a
  limited parameter space and thus the transition is found to be random
  (Type-III).
\end{enumerate}

\section{Proofs} \label{appendix}

In this section, we give the proofs of our main theorems. Our strategy is first
to obtain the reduction onto the center manifold of the dynamics and second, to
analyze the reduced equations.

\subsection{Center manifold approximation}
We recall that the center part of the solution is given by
\[
  u_c = u_1(t) f_1 + u_2(t) f_2
\]
The rest of the solution is approximated by the center manifold which we expand
as
\begin{equation}
  \label{eq:cm-derive1}
  \Phi = \sum_{k \ge 3} \Phi_k(t) f_k
\end{equation}
As is well known, see \cite{ptd}, the lowest order (quadratic) approximation of $\Phi$ is
obtained as the solution of
\begin{equation}
  \label{eq:cm-derive2}
  \mathcal{L} \Phi = -P_s G(u_c) + o(2).
\end{equation}
Here $P_s:X \to E_s$ is the projection operator where $E_s$ is the stable space
which is the span of $\{ f_j : j \ge 3 \}$ in $X$ and $\mathcal{L} = L
\mid_{E_s}$ is the restriction of the linear operator onto the stable space.

Since $f_j \in E_s$, for $j \ge 3$, we have
\begin{equation}
  \label{eq:cm-derive3}
  \langle P_s (\cdot), f_j^* \rangle = \langle \cdot, f_j^* \rangle, \qquad j \ge 3.
\end{equation}
Plugging \eqref{eq:cm-derive1} in \eqref{eq:cm-derive2}, taking the inner product
of \eqref{eq:cm-derive2} with $f_j$, $j \ge3$ and making use of
\eqref{eq:cm-derive3}, we get
\begin{equation}
  \label{eq:cm-derive4}
  \Phi_j \beta_j \langle f_j, f_j^* \rangle = \langle \Phi, L^* f_j \rangle =
  -\langle G(u_c), f_j^* \rangle, \qquad j \ge 3.
\end{equation}
From \eqref{eq:cm-derive4}, using the notation \eqref{non-not}, we obtain the
below formula for the coefficients of the center manifold.
\begin{equation}
  \label{eq:cm-deriv5}
  \begin{aligned}
    \Phi_j & = \frac{-1}{\beta_j \langle f_j, f_j^* \rangle} \langle G(u_c),
    f_j^{\ast} \rangle \\ 
    & = \frac{-1}{\beta_j} \sum_{m, n = 1, 2} u_m u_n G_2(m,n,j), \quad j \ge 3.
  \end{aligned}
\end{equation}
Notice that $\Phi_j$ is independent of the trilinear operator $G_3$.

Now the reduced equations of the system are obtained by plugging
\[
  u = u_c + \Phi
\]
into the main equation \eqref{main}, which is basically considering the dynamics
on the center manifold and taking projection onto the center space, that is the
span of the $\{ f_1, f_2 \}$. This gives
\[
  \frac{d}{dt} \langle u, f_k^* \rangle =
  \langle L_{\lambda} u, f_k^* \rangle + \langle G_{\lambda}(u), f_k^*
  \rangle, \qquad k =1, 2,
\]
which is equivalent to the system
\begin{equation}\label{pre-red}
  \frac{du_k}{dt} =  \beta_k u_k + \frac{1}{\langle f_k, f_k^* \rangle }
  \langle G(u_c+\Phi), f_k^{\ast} \rangle, \qquad k=1,2.
\end{equation}
To obtain a closed system, we have to write the nonlinear term above as a
function of $u_1$, $u_2$. This can be obtained as follows
\[
  \begin{aligned}
    & \frac{1}{\langle f_k, f_k^* \rangle }
    \langle G(u_c+\Phi), f_k^{\ast} \rangle \\
    & = \frac{1}{\langle f_k, f_k^* \rangle }
    \langle G_2(u_c, u_c) + G_2^s(u_c, \Phi) + G_3(u_c, u_c, u_c), f_k^{\ast} \rangle + O(4) \\
    & := P_{2, k} + P_{3, k} + O(4), \qquad k=1,2 \\
  \end{aligned}
\]
and \eqref{pre-red} becomes
\begin{equation}\label{pre-red1}
  \frac{du_k}{dt} =  \beta_k u_k + P_{2, k}(u_1, u_2) + P_{3, k}(u_1, u_2) +
  O(4), \qquad k=1,2.
\end{equation}
where $P_{2, k}(u_1, u_2)$ denote the quadratic terms in $u_1$, $u_2$ given by
\[
  \begin{aligned}
    P_{2, k} & =
    \frac{1}{\langle f_{k}, f_{k}^* \rangle }\langle G_2(u_c, u_c), f_k^{\ast}
    \rangle \\
    & = \frac{1}{\langle f_{k}, f_{k}^* \rangle }
    \sum_{i,j=1}^2 \langle G_2(u_i f_i, u_j f_j), f_k^{\ast} \rangle \\
    & = \sum_{i,j=1}^2 u_i u_j G_2(i,j,k), \qquad k=1,2
  \end{aligned}
\]
and $P_{3, k}(u_1, u_2)$, $k=1,2$ denote the cubic terms given by
\[
  \begin{aligned}
    P_{3, k} & =
    \frac{1}{\langle f_{k}, f_{k}^* \rangle }
    \langle G_2^s(u_c, \Phi) + G_3(u_c, u_c, u_c), f_k^{\ast} \rangle \\
    & = \frac{1}{\langle f_{k}, f_{k}^* \rangle }
    \sum_{i, l, m \in \left\{ 1,2 \right\}, j\ge3}^{}
    \langle G_2^s(u_i f_i, \Phi_j f_j) + G_3(u_i f_i, u_m f_m, u_n f_n) , f_k^{\ast} \rangle \\
    & = \sum_{m,n \in \{1,2\},  j \ge 3} 
    u_i u_m u_n   \frac{-1}{\beta_j} G_2(m,n,j) G_2^s(i,j,k) \\
    & \qquad + \sum_{i,m,n \in \{1,2\} } u_i u_m u_n  G_3(i,m,n,k) 
  \end{aligned}
\]
Now, we take into account the structure \eqref{eigenmode-assumption} of the
eigenmodes into account.
\subsection{Consequences of assumption on nonlinear operator}

The condition \eqref{main-nonlinear-assumption} has the following implications.
\begin{align}
  & (l_1, l_2) \notin C_{11} \implies \langle
    (G_2(e_{j_1, j_2}, e_{j_1, j_2}), e_{l_1, l_2}) \rangle = 0 \label{a3} \\
  & (l_1, l_2) \notin C_{12} \implies \langle
    (G_2(e_{j_1, j_2}, e_{k_1, 0}),  e_{l_1, l_2}) \rangle = 0 \label{a1} \\
  & (l_1, l_2) \notin C_{12} \implies \langle
    (G_2(e_{k_1, 0}, e_{j_1, j_2}),  e_{l_1, l_2}) \rangle = 0 \label{a2} \\
  & (l_1, l_2) \notin C_{22} \implies \langle
    (G_2(e_{k_1, 0}, e_{k_1, 0}),   e_{l_1, l_2}) \rangle = 0 \label{a4}
\end{align}
where the index sets are as defined in \eqref{Cij}. 

Using the notation \eqref{non-not}, the observations in \eqref{a3}--\eqref{a4}
lead to the following observations
\begin{equation}\label{ortho-1}
  G_2(1,1,1) = G_2(2,2,1) = 0,
\end{equation}
\begin{equation}\label{ortho-2}
  G_2(2,2,2) = G_2(2,1,2) = G_2(1,2,2) = 0.
\end{equation}
But it is possible that
\begin{equation}\label{non-ortho}
  G_2(1,2, 1) \ne 0, \qquad G_2(2,1,1) \ne 0, \qquad G_2(1,1,2) \ne 0
\end{equation}

By \eqref{Cij}, we note that
\begin{displaymath}
  C_{11} \cap C_{12} = C_{12} \cap C_{22} = \emptyset
\end{displaymath}
which implies that the following products of nonlinear interactions of the
critical modes with higher modes vanish.
\begin{equation}\label{ortho-3}
  \begin{aligned}
    & G_2(\sigma_1(1,1,j))G_2(\sigma_2(1,2,j)) = 0, \quad \forall j \ge 3, \text{
      since } C_{11} \cap C_{12} = \emptyset \\
    & G_2(\sigma_1(1,2,j))G_2(\sigma_2(2,2,j)) = 0, \quad \forall j \ge 3, \text{
      since } C_{12} \cap C_{22} = \emptyset
  \end{aligned}
\end{equation}
where $\sigma_1$, $\sigma_2$ are any permutations of the set $\{ 1,2,3 \}$. For example, the first condition above is equivalent to the vanishing of the 18 products given by
\[
  \begin{aligned}
    & G_2(1,1,j)G_2(1,2,j) = G_2(1,j,1)G_2(1,2,j) = G_2(j,1,1)G_2(1,2,j) = 0,  \\
    & G_2(1,1,j)G_2(1,j,2) = G_2(1,j,1)G_2(1,j,2) = G_2(j,1,1)G_2(1,j,2) = 0, \\
    & \qquad  \vdots
  \end{aligned}
\]
and the same for the remaining $12$ products as well for all $j \ge 3$.

Similarly, for the trilinear term, for $i, j, k, l \in \{1, 2\}$,
\begin{equation}
  \label{eq:tri1}
  i+j+k+l = 1 \mod 2 \implies G_3(i, j, k, l) = 0 \implies G_3^s(i, j, k, l) = 0.
\end{equation}

\subsection{Structure of the quadratic polynomials $P_{2,k}$}
As a consequence of \eqref{ortho-2}
\begin{equation}\label{ortho-c4}
  G_2^s(1,2,2) = G(1,2,2) + G(2,1,2) = 0.
\end{equation}
Hence by \eqref{ortho-1}, \eqref{ortho-2}, and \eqref{ortho-c4}, the quadratic
terms become
\begin{equation} \label{eq:P2-1}
   \begin{aligned}
    P_{2, 1} & = u_1^2 G(1,1,1) + u_1 u_2 G_2^s(1, 2, 1) + u_2^2 G(2,2,1) \\
    & = u_1 u_2 G_2^s(1,2,1) 
  \end{aligned}
\end{equation}
\begin{equation} \label{eq:P2-2}
  \begin{aligned}
    P_{2, 2} & = u_1^2 G(1, 1, 2) + u_1 u_2 G_2^s(1, 2, 2) + u_2^2 G(2,2,2) \\
    & = u_1^2 G(1, 1, 2)
  \end{aligned}
\end{equation}

\subsection{Structure of the cubic polynomials $P_{3,k}$}
The coefficients of the following cubic terms vanish by \eqref{ortho-3} and
\eqref{eq:tri1}.




\begin{enumerate}
  \item The term $u_1^2 u_2$ in $P_{3, 1}$
  \[
    \begin{aligned}
      G_3^s(1,1,2,1) + \sum_{j \ge 3} \frac{-1}{\beta_j} &
      ( G(1,2,j) G_2^s(1,j,1) + G(2,1,j) G_2^s(1,j,1) \\
      & + G(1,1,j) G_2^s(2,j,1) ) = 0,
    \end{aligned}
  \]
  \item The term $u_2^3$ in $P_{3, 1}$
  \[
    G_3(2, 2, 2, 1) + \sum_{j\ge3} \frac{-1}{\beta_j} G(2,2,j) G_2^s(2, j, 1) = 0,
  \]
  \item The term $u_1 u_2^2$ in $P_{3, 2}$
  \[
    \begin{aligned}
      G_3^s(1, 2, 2, 2) + \sum_{j \ge 3} \frac{-1}{\beta_j} &
      ( G(1,2,j) G_2^s(2,j,2) + G(2,1,j) G_2^s(2,j,2) \\
      & + G(2,2,j) G_2^s(1,j,2) ) +  = 0,
    \end{aligned}
  \]
  \item The term $u_1^3$ in $P_{3, 2}$
  \[
    G_3(1, 1, 1, 2) + \sum_{j\ge3} \frac{-1}{\beta_j} \left(
      G(1,1,j)G_2^s(1,j,2) \right) = 0,
  \]
\end{enumerate}

\subsection{The existence of bifurcated steady state solutions}

We start by finding the straight line orbits of the vector field $(F_1,F_2)$
given by \eqref{F1F2} near $(u_1, u_2) = (0, 0)$. Clearly $u_1=0$ is always a
straight line orbit that is $du_1/dt = 0$. On $u_1 = 0$, the dynamics of $u_2$
is as follows.
\begin{equation} \label{strline2}
  \frac{du_2}{dt}= u_2 (\beta +b_3 u_2^2) + o(u_2^4). 
\end{equation}
For $\beta b_3 < 0$, we find the two bifurcated steady state solutions (roll
pattern solutions) $R_1$, $R_2$ given by \eqref{steady-roll}, that is with
amplitudes $u_1=0$, and
\begin{equation} \label{roll-amplitudes}
    u_1 = 0, \qquad u_2 = \pm \sqrt{-\frac{\beta}{b_3}} + o(\sqrt{-\beta}).
\end{equation}

Now we look for other straight line orbits $u_1=k u_2$, $k \ne 0$ near the
origin. On such a straight line orbit, the following relation is satisfied
\[
  k=\frac{u_1}{u_2} = \frac{F_1}{F_2} =
  \frac{a_1 k u_2^2 + O(u_2^3)}{b_1 k^2 u_2^2 +O(u_2^3)}, \qquad u_2 \to 0.
\]
Hence, for $k \ne 0$, $k$ should satisfy the relation
\[
  k^2 b_1 = a_1.
\]
In particular, if $a_1 b_1<0$, no such $k \ne 0$ exists.

On the other hand for $a_1b_1>0$, in addition to $u_1=0$, four more straight
line orbits appear
\[
  u_1= k_i u_2, \qquad k_i=(-1)^{i}\sqrt{a_1/b_1}, \quad i=1,2.
\]
The flow on the straight line $u_1=k_i u_2$ is given by
\begin{equation}\label{strline}
  \frac{du_2}{dt}=u_2(\beta+b_1 k_i^2 u_2+(b_2 k_i^2+ b_3)u_2^2),
\end{equation}
which has two steady state solutions
\begin{equation} \label{hex-amp}
    \begin{aligned}
    & u_1 = k_i u_2 = (-1)^{i+1}\frac{\beta}{\sqrt{a_1 b_1}} + O(\beta^2), \\
    & u_2 = -\frac{\beta}{b_1k_i^2} + O(\beta^2) = -\frac{\beta}{a_1} +
    O(\beta^2), \qquad i=1,2.
  \end{aligned}
\end{equation}
On each line $u_1 = k_i u_2$, $i=1,2$, there is a bifurcated steady state
solution which gives the amplitudes of (hexagonal pattern solution) $H_1$ and
$H_2$ given by \eqref{steady-hex}.

\subsection{The stability of the bifurcated steady state solutions}



The Jacobian matrix of the right hand side of the reduced equations
\eqref{reduced} be
\begin{equation} \label{jacobian}
  DF(u_1,u_2)=
  \left(
    \begin{array}{cc}
      3 a_2 u_1^2+a_3 u_2^2+a_1 u_2+\beta  & u_1 \left(2 a_3 u_2+a_1\right) \\
      2 u_1 \left(b_2 u_2+b_1\right) & b_2 u_1^2+3 b_3 u_2^2+\beta  \\
    \end{array}
  \right).
\end{equation}
Let us also denote the eigenvalues of $DF$ by $\lambda_1$, $\lambda_2$. We find
that the eigenvalues of $DF$ at the steady states \eqref{roll-amplitudes} as
below. For $R_1$, that is for $(u_1, u_2) = (0, -\sqrt{-\frac{\beta}{b_3}}) +
o(\sqrt{-\beta})$, they are
\begin{displaymath}
      \lambda_1 = -2 \beta, \quad \lambda_2 = -a_1 \sqrt{-\frac{\beta }{b_3}} + O(\beta),
\end{displaymath}
and for $R_2$, that is for $(u_1, u_2) = (0, \sqrt{-\frac{\beta}{b_3}}) +
o(\sqrt{-\beta})$, they are
\begin{displaymath}
  \lambda_1 = -2 \beta, \quad \lambda_2 = a_1 \sqrt{-\frac{\beta }{b_3}} + O(\beta).
\end{displaymath}
Thus the stability of the bifurcated solutions $R_1$, $R_2$ given by
\eqref{steady-roll} for $\beta> 0$ are as follows.
\[
  \begin{aligned}
    a_1 > 0 \qquad & \text{$R_1$ is a stable node, $R_2$ is a saddle} \\
    a_1 < 0 \qquad & \text{$R_1$ is a saddle, $R_2$ is a stable node} \\
  \end{aligned}
\]


Since the amplitude \eqref{hex-amp} of the mixed mode steady states are $u_1 =
(-1)^{i+1} \frac{ \beta }{ \sqrt{ a_1 b_1 } } + O(\beta^2)$, $u_2 = - \frac{
\beta }{ a_1 } + O(\beta^2)$, we find that the Jacobian matrix of the mixed
modes is
\begin{displaymath}
  \begin{pmatrix}
    \beta+a_1 u_2 + O(\beta^2)& a_1 u_1 + O(\beta^2) \\
    2b_1 u_1 + O(\beta^2) & \beta + O(\beta^2)
  \end{pmatrix}
\end{displaymath}
and the corresponding eigenvalues satisfy
\begin{displaymath}
  \lambda_1 + \lambda_2 = \beta + O(\beta^2)
\end{displaymath}
and
\begin{displaymath}
  \lambda_1 \lambda_2 = -2 a_1 b_1 u_1 u_2 + O(\beta^3) = -2 \beta^2 + O(\beta^3)
\end{displaymath}
Thus we find the eigenvalues and the corresponding eigenvectors of the
amplitudes of the equilibria $H_i$ as 
\[
  H_1:
  \quad \lambda_1 = -\beta + O(\beta^2), v_1 = 
  \begin{pmatrix}
    - \frac{ a_1 }{ \sqrt{ a_1 b_1 } } \\ 1
  \end{pmatrix},
  \quad \lambda_2 = 2 \beta + O(\beta^2), v_2 = 
  \begin{pmatrix}
    \frac{ a_1 }{ 2 \sqrt{ a_1 b_1 } } \\ 1
  \end{pmatrix}
\]
and
\[
  H_2:
  \quad \lambda_1 = -\beta + O(\beta^2), v_1 = 
  \begin{pmatrix}
    \frac{ a_1 }{ \sqrt{ a_1 b_1 } } \\ 1
  \end{pmatrix},
  \quad \lambda_2 = 2 \beta + O(\beta^2), v_2 = 
  \begin{pmatrix}
    - \frac{ a_1 }{ 2 \sqrt{ a_1 b_1 } } \\ 1
  \end{pmatrix}
\]

Since the eigenvalues have always opposite signs near $\beta = 0$, the
equilibria $H_i$ given by \eqref{steady-hex} are always saddles whenever they exist.

The details of the proof of the main theorems follow from the above analysis and
the dynamic transition theory \cite{ptd, ma2005}.

\section{Summary and discussion}
In this paper, we have found all possible transition scenarios of a general
dissipative system with two dimensional critical center space. We have two main
assumptions. First is the one that the one of the eigenmodes has a rectangle
pattern and the other one has a roll pattern which is possible in spatial
domains of at least two dimensions with homogeneous boundary conditions. The
second one is an orthogonality condition on the nonlinear operator with respect
to the basis vectors. This condition is generally satisfied when the bilinear
and trilinear terms of the Taylor expansion of the nonlinear operator is a
product of the unknown function and its derivatives.

Under the above general conditions, we manage to show that all three types of
transitions of the dynamic transition theory are possible at the first
criticality $\lambda_c$ under generic conditions. We also determine the
bifurcated steady state solutions with their stability and the bifurcated
attractors in each case. Finally, we give several applications to demonstrate
that all the conclusions of our main theorems are observable.

Up until now, the dynamic transition theory has been used as a general tool to
understand specific problems. This paper points in the direction of
determination of transitions and pattern formations for a general class of
problems. Thus we believe that the results presented in this study will provide
a general framework for the determination and the validity of transitions in
many applications.

\bibliographystyle{plain}
\bibliography{main}

\end{document}